\documentstyle[12pt]{article}
\author{ V.\ A.\ SLOBODENYUK}
\title{ON THE EVOLUTION OPERATOR KERNEL FOR THE COULOMB AND
COULOMB--LIKE POTENTIALS}
\date{}

\begin{document}

\maketitle

\begin{abstract}
With a help of the Schwinger --- DeWitt expansion analytical
properties of the evolution operator kernel for
the Schr\"odinger equation in time variable $t$ are studied
for the Coulomb and Coulomb-like (which behaves themselves
as $1/|\vec q|$ when $|\vec q| \to 0$) potentials.
It turned out to be that the Schwinger --- DeWitt
expansion for them is divergent. So, the kernels for these
potentials have additional (beyond $\delta$-like)
singularity at $t=0$. Hence, the initial condition is fulfilled
only in asymptotic sense. It is established that the potentials
considered do not belong to the class of potentials, which
have at $t=0$ exactly $\delta$-like singularity and for
which the initial condition is fulfilled in rigorous sense
(such as $V(q) = -\frac{\lambda (\lambda-1)}{2} \frac {1}{\cosh^2 q}$
for integer $\lambda$).
\end{abstract}

\newpage

\bigskip

\section{Introduction}

This paper continues the series of works~\cite{S1,S2,S3} devoted to
study of dependence of the evolution operator kernel for the
Schr\"odinger equation on time interval $t$ (especially, in
vicinity of origin). We use for the kernel the Schwinger --- DeWitt
expansion~\cite{Sch,DeW1,DeW2} which in one-dimensional case reads
  \begin{equation} \label{f1.1}
  \langle q',t\mid q,0 \rangle = \frac{1}{\sqrt{2\pi it}}
  \exp \left\{i \frac{(q'-q)^2}{2t} \right\} F(t;q',q),
  \end{equation}
where
  \begin{equation} \label{f1.2}
  F(t;q',q) = \sum_{n=0}^{\infty} (it)^n a_n(q',q).
  \end{equation}

It was obtained~\cite{S2} the estimate for the coefficients $a_n$
which shows that this expansion is usually divergent, if there is
no any cancellations of different contributions. Such cancellations
really take place for some potentials, as it is established
in~\cite{S3}\,. For example, for the potential
  \begin{equation} \label{f1.3}
  V(q) = -\frac{\lambda (\lambda-1)}{2} \frac {1}{\cosh^2 q}
  \end{equation}
the series~(\ref{f1.2}) converges when $\lambda$ is integer. Thus,
for the most of the potentials the expansion~(\ref{f1.2}) is
divergent, but there exist the class of potentials for which this
expansion
is convergent at some discrete values of the coupling constant~$g$.

Divergence of expansion~(\ref{f1.2}) shows that the function $F$,
which is really
  \begin{equation} \label{f1.4}
  F(t;q',q)=\frac{\langle q',t\mid q,0 \rangle_V}
                 {\langle q',t\mid q,0 \rangle_{V=0}}
  \end{equation}
(subscribe $V$ means that the kernel is taken for the potential $V$),
has essential singularity at the point $t=0$. Hence, $F$ has no any
meaning at this point and $F$ tends to 1 for $t \to 0$ only in
asymptotic sense.

Singularity of $\langle q',t\mid q,0 \rangle_{V=0}$ at $t=0$ is
acceptable and necessary, because
  \begin{equation} \label{f1.5}
  \langle q',t=0\mid q,0 \rangle_{V=0} = \delta (q'-q)
  \end{equation}
in correspondence with initial condition for the kernel. But if
$F$ has additional singularity (namely this singularity is
under consideration here), then one cannot say that
  \begin{equation} \label{f1.6}
  \langle q',t=0\mid q,0 \rangle_{V} =
           \langle q',t=0\mid q,0 \rangle_{V=0} = \delta (q'-q).
  \end{equation}
One can say only that
  \begin{equation} \label{f1.7}
  \langle q',t \to +0\mid q,0 \rangle_{V} \to
           \langle q',t=0\mid q,0 \rangle_{V=0} = \delta (q'-q)
  \end{equation}
when $t$ is real positive. Relation~(\ref{f1.7}) has asymptotic
character. Using it one can determine unambiguously evolution
of the system, but asymptotic initial condition leads to
appearance of divergences in different expressions. If one
works with the potentials, for which the Schwinger --- DeWitt
expansion converges, e.g., with~(\ref{f1.3}), then problem
of divergence does not arise at all.
One may assume in this connection that such
potentials have emphasized meaning in quantum theory. This is
why it is interesting to probe convergence of the
expansion~(\ref{f1.2}) for different frequently used potentials.

In present paper the Coulomb and Coulomb-like (the potentials
which behave themselves as $1/|\vec q|$ when $|\vec q| \to 0$
and are regular in some vicinity of the point $|\vec q|=0$)
potentials are under consideration.
In the case of spherically symmetric potentials the three-dimensional
kernel can be reduced to infinite sequence of one-dimensional ones
for the effective potentials. For these kernels the technique
developed in~\cite{S3} is used.

\section{The Schwinger --- DeWitt expansion for spherically
symmetric potentials}

Let us consider the Schr\"odinger equation for the evolution
operator kernel in three-dimensional space
  \begin{equation} \label{f1}
  i \frac{\partial}{\partial t} \langle \vec q\,',t\mid \vec q,0 \rangle =
  -\frac{1}{2} \sum_{i=1}^3 \frac{\partial^2}{\partial q'^2_i}
  \langle \vec q\,',t\mid \vec q,0 \rangle + V(\vec q\,')
  \langle \vec q\,',t\mid \vec q,0 \rangle
  \end{equation}
with initial condition
  \begin{equation} \label{f2}
  \langle \vec q\,',t=0\mid \vec q,0 \rangle = \delta (\vec q\,'-\vec q).
  \end{equation}
Here and everywhere dimensionless values are used. If the potential
$V(\vec q)$ depends on $q=|\vec q|$ only then it is more convenient
to transfer to radial equation by means of representation
  \begin{equation} \label{f3}
  \langle \vec q\,',t\mid \vec q,0 \rangle =
  \sum_{l=0}^{\infty} \frac{2l+1}{4\pi q' q} P_l(\cos \gamma)
  \langle q',t\mid q,0 \rangle_l,
  \end{equation}
where $\gamma$ is the angle between the vectors $\vec q\,'$ and
$\vec q$, $P_l$ are the Legendre polynomials,
$\langle q',t\mid q,0 \rangle_l$ is "one-dimensional kernel"
depending on absolute values of the vectors $\vec q\,'$,
$\vec q$ and on integer number $l$. The variables $q'$, $q$
vary at the positive half-line.

Substitution of~(\ref{f3}) into~(\ref{f1}) gives equation
  \begin{equation} \label{f4}
  i \frac{\partial}{\partial t} \langle q',t\mid q,0 \rangle_l =
  -\frac{1}{2} \frac{\partial^2}{\partial q'^2}
  \langle q',t\mid q,0 \rangle_l +
  \left( \frac{l(l+1)}{2}\frac{1}{q'^2} + V(q')\right)
   \langle q',t\mid q,0 \rangle_l,
  \end{equation}
which coincides in form with the one-dimensional Schr\"odinger
equation for effective potential
  \begin{equation} \label{f5}
  U_l(q)= \frac{l(l+1)}{2}\frac{1}{q^2} + V(q).
  \end{equation}
The initial condition, as it is clear from consideration of the
problem for $V(q)\equiv 0$, is to be taken in the form
  \begin{equation} \label{f6}
  \langle q',t=0\mid q,0 \rangle_l = \delta (q'-q)-(-1)^l\delta(q'+q),
  \end{equation}
which provides correct behaviour of the solution at $q=0$.

According to general procedure~\cite{DeW1,DeW2,S3} we represent
  \begin{eqnarray} \label{f7}
  \langle q',t\mid q,0 \rangle_l &=& \frac{1}{\sqrt{2\pi it}}
  \exp \left\{i \frac{(q'-q)^2}{2t} \right\} F_l^{(-)}(t;q',q)-
  \nonumber \\ &&
  (-1)^l \frac{1}{\sqrt{2\pi it}}
  \exp \left\{i \frac{(q'+q)^2}{2t} \right\} F_l^{(+)}(t;q',q).
  \end{eqnarray}
The equations for the functions $F_l^{(\pm)}$ are
  \begin{equation} \label{f8}
  i \frac{\partial F_l^{(\pm)}}{\partial t} =
  -\frac{1}{2} \frac{\partial ^2 F_l^{(\pm)}}{\partial q'^2} +
  \frac{q'\pm q}{it} \frac{\partial F_l^{(\pm)}}{\partial q'}
  + U_l(q')F_l^{(\pm)}.
  \end{equation}
Initial conditions are
  \begin{equation} \label{f9}
  F_l^{(\pm)}(t=0;q',q)=1.
  \end{equation}
Here $F_l^{(\pm)}$ are defined for $q',\; q >0$. However, one
can consider the analytical continuation into domain $q<0$. It
is necessary for this to come out into the complex plain of the
variable $q$, because continuation along the real axis cannot be
done so as the potential $U_l(q)$ and, hence, the functions
$F_l^{(\pm)}$ have singularity at $q=0$. It is clear from~(\ref{f8}),
(\ref{f9}) that $F_l^{(+)}(t;q',q)=F_l^{(-)}(t;q',-q)$ ($q>0$).
So, instead of two functions $F_l^{(+)}$ and $F_l^{(-)}$ one may
consider only one function $F_l=F_l^{(-)}$, which is defined for
$q$ varying along the hole real line. Instead of two
equations~(\ref{f8}) it is enough to consider only one equation
for the function $F_l$ taking at~(\ref{f8}) the lower sign.

Let us look for $F_l$ in the form
  \begin{equation} \label{f10}
  F_l(t;q',q) = \sum_{n=0}^{\infty} (it)^n a_n^l(q',q).
  \end{equation}
For the coefficient functions following relations are obtained
  \begin{equation} \label{f11}
  a_0^l(q',q)=1,
  \end{equation}
  \begin{equation} \label{f12}
  na_n^l + (q'-q) \frac{\partial a_n^l}{\partial q'}
  = \frac{1}{2} \frac{\partial ^2 a_{n-1}^l}{\partial q'^2} -
  U_l(q')a_{n-1}^l.
  \end{equation}
The solutions of these equations can be represented as~\cite{S1}
  \begin{equation} \label{f13}
  a_n^l(q',q)= \int \limits_0^1 \eta^{n-1} d\eta
  \left(\frac{1}{2} \frac{\partial ^2}{\partial \tilde q^2} -
  U_l(\tilde q)\right) a_{n-1}^l(\tilde q,q)\Biggl.
  \Biggr|_{\tilde q=q+(q'-q)\eta}.
  \end{equation}
Another representation for $a_n^l$ can be derived starting from the
expansion for the potential
  \begin{equation} \label{f14}
  U_l(q')= \sum_{k=0}^{\infty} (q'-q)^k \frac{U_l^{(k)}(q)}{k!},
  \end{equation}
where
  $$U_l^{(k)}(q) \equiv \frac{d^k U_l(q)}{dq^k}.$$

It is possible to write for such $q', \; q$, for which
equation~(\ref{f14}) takes place,
  \begin{equation} \label{f15}
  a_n^l(q',q)= \sum_{k=0}^{\infty} (q'-q)^k b_{nk}^l(q).
  \end{equation}
Then $b_{00}^l(q)=1$ and for $n>0$ one has recurrent relations
  \begin{equation} \label{f16}
  b_{nk}^l=\frac{1}{n+k} \left[\frac{(k+1)(k+2)}{2} b_{n-1, k+2}^l -
  \sum_{m=0}^k \frac{U_l^{(m)}(q)}{m!} b_{n-1, k-m} \right]
  \end{equation}
Note, that because of singularity of $U_l(q)$ at $q=0$ the
expansion~(\ref{f14}) is not valid for $q<0$. So,
equations~(\ref{f16}) can be used directly for calculation 
of $b_{nk}^l(q)$
only in the domain $q>0$. However, for all potentials considered
at present paper the expansion~(\ref{f10}) is divergent. To prove
divergence it is enough to show that only for one of two functions
$F_l^{(+)}$ or $F_l^{(-)}$ and for one number
$l$ the series of type~(\ref{f10}) diverges. Therefore we
will consider only positive $q$.

The relations obtained will be used for analysis of convergence of
the Schwinger --- DeWitt expansion for the Coulomb and Coulomb-like
potentials.

\section{The Coulomb potential}

Let us consider the Coulomb potential
  \begin{equation} \label{f17}
  V(q)=\frac{\alpha}{q}.
  \end{equation}
The effective potential is
  \begin{equation} \label{f18}
  U_l(q)=\frac{l(l+1)}{2} \frac{1}{q^2} + \frac{\alpha}{q}.
  \end{equation}
"One-dimensional kernel" $\langle q',t\mid q,0 \rangle_l$ is
the coefficient at the expansion~(\ref{f3}) for the initial
three-dimensional kernel $\langle \vec q\,',t\mid \vec q,0 \rangle$
in the Legendre polynomials $P_l$. Therefore, if the
expansion~(\ref{f10}) in powers of $t$ for
$\langle q',t\mid q,0 \rangle_l$ is divergent at least for any one
value of $l$, then the Schwinger --- DeWitt expansion for
$\langle \vec q\,',t\mid \vec q,0 \rangle$ is divergent too. We
take $l=0$. Then $U_0(q)=V(q)=\alpha/q$.

With a help of representation~(\ref{f13}) one can consequently
calculate the coefficient functions $a_n^0$. For example,
  \begin{equation} \label{f19}
  a_1^0(q',q)= -\frac{\alpha}{q'-q} \log \frac{q'}{q},
  \end{equation}
  \begin{equation} \label{f20}
  a_2^0(q',q)=\frac{\alpha^2}{2} \left( \frac{1}{q'-q}
  \log \frac{q'}{q}\right)^2 +
  \frac{\alpha}{(q'-q)^3} \log \frac{q'}{q} -
  \frac{\alpha}{2}\frac{1}{(q'-q)^2}
  \left( \frac{1}{q'} + \frac{1}{q} \right),
  \end{equation}
etc.

Nevertheless, to evaluate behaviour of $a_n^0$ for
$n \to \infty$ it is more convenient to use~(\ref{f15}),
(\ref{f16}). Calculate derivatives
  \begin{equation} \label{f21}
  V^{(m)}(q)=\frac{(-1)^m \alpha m!}{q^{m+1}}.
  \end{equation}
By means of~(\ref{f16}) we find
  \begin{equation} \label{f22}
  b_{1k}^0=\frac{(-1)^{k+1}}{k+1} \frac{\alpha}{q^{k+1}},
  \end{equation}
  \begin{equation} \label{f23}
  b_{2k}^0=\frac{(-1)^{k+1}}{k+2} \left[
  \frac{(k+1)(k+2)}{2(k+3)} \frac{\alpha}{q^{k+3}} -
   \sum_{m=0}^k \frac{1}{m+1} \frac{\alpha^2}{q^{k+2}} \right],
  \end{equation}
  \begin{eqnarray} \label{f24}
  b_{3k}^0&=&\frac{(-1)^{k+1}}{k+3} \left[
  \frac{(k+1)(k+2)(k+3)(k+4)}{2^2(k+4)(k+5)} \frac{\alpha}{q^{k+5}} -
    \right. \nonumber \\ &&
  \frac{1}{2}  \left(
  \frac{(k+1)(k+2)}{k+4} \sum_{m=0}^{k+2} \frac{1}{m+1}  +
  \sum_{m=0}^k \frac{m+1}{m+3} \right) \frac{\alpha^2}{q^{k+4}} +
    \nonumber \\ &&\left.
  \sum_{m=0}^k \frac{1}{m+2}
  \sum_{m'=0}^m \frac{1}{m'+1} \frac{\alpha^3}{q^{k+3}} \right].
  \end{eqnarray}

It is easy to understand that the coefficients $b_{nk}^0$ in this
case have the following structure
  \begin{equation} \label{f25}
  b_{nk}^0(q)= \sum_{j=1}^n (-1)^{k+j} C_{nk}^j
  \frac{\alpha^j}{q^{k+2n-j}},
  \end{equation}
and besides the numerical coefficients $C_{nk}^j$ and all partial
contributions into them are positive. It means that for every $m$
all contributions into coefficients in front of $1/q^m$ has the
same sign and nothing cancellations do occur. In this case, as
it follows from the analysis of representation~(\ref{f13})
\cite{S2}, the coefficients $b_{nk}^0$ and $a_n^0$ will increase
as $n!$ at $n \to \infty$.

One can convinced of this directly from relations~(\ref{f16}). Put
$k=0$ and calculate $C_{n0}^1$. It is obviously, that
  \begin{equation} \label{f26}
  C_{n0}^1 = \frac{(n-1)!}{2^{n-1}(2n-1)} \sim n!
  \end{equation}
Because the value $-C_{n0}^1$ is the coefficient in front of independent
structure $(it)^n \Delta q^0/ q^{2n+1}$ at the
expansion~(\ref{f10}), then its factorial growth means that
$|a_n| \sim n!$ for $n \to \infty$.

Thus we established divergence of the Schwinger --- DeWitt
expansion for the Coulomb potential. This means that the initial
condition~(\ref{f2}) (or~(\ref{f6})) for the kernel cannot
be fulfilled because of essential singularity of the solution
of the Schr\"odinger equation at the point $t=0$. So,
the evolution operator kernel for the
Coulomb potential exists only in asymptotic sense.

\section{Other potentials with Coulomb-like singularity $1/|\vec q|$}

Consider spherically symmetric potential of the form
  \begin{equation} \label{f4.1}
  V(q)=\frac{\alpha}{q} +f(q),
  \end{equation}
where $f(q)$ is regular in some vicinity of zero function, which
can be represented at this domain by the convergent series
  \begin{equation} \label{f4.2}
  f(q)= \sum_{k=0}^{\infty} f_k q^k.
  \end{equation}
For example, the Yukawa potential
  \begin{equation} \label{f4.3}
  V(q)=\frac{\alpha}{q} e^{-\beta q}
  \end{equation}
belongs to this class.

Analogously to previous section, to analyse the Schwinger ---
DeWitt expansion for the kernel
$\langle \vec q\,',t\mid \vec q,0 \rangle$ for convergence we
consider "one-dimensional kernel"
$\langle q',t\mid q,0 \rangle_l$ for $l=0$. Then the effective
potential is $U_0(q)=V(q)$. The derivatives of the potential
read
  \begin{equation} \label{f4.4}
  U_0^{(m)}(q)=V^{(m)}(q)= \frac{(-1)^m \alpha m!}{q^{m+1}}+
  \sum_{l=m}^{\infty} \frac{l!}{(l-m)!} f_l q^{l-m}.
  \end{equation}

One can consequently determine all $b_{nk}^0$ from
relations~(\ref{f16}). For instance,
  \begin{equation} \label{f4.5}
  b_{1k}^0=\frac{(-1)^{k+1}}{k+1} \frac{\alpha}{q^{k+1}} -
  \frac{1}{k+1} \sum_{l=k}^{\infty}
  \frac{l!}{k!(l-k)!} f_l q^{l-k},
  \end{equation}
  \begin{eqnarray} \label{f4.6}
  b_{2k}^0 &=& \frac{(-1)^{k+1}}{k+2} \left[
  \frac{(k+1)(k+2)}{2(k+3)} \frac{\alpha}{q^{k+3}} -
   \sum_{m=0}^k \frac{1}{m+1} \frac{\alpha^2}{q^{k+2}} \right] +
    \nonumber \\ &&
  \frac{1}{k+2} \left[ -\frac{(k+1)(k+2)}{2(k+3)}
  \sum_{l=k+2}^{\infty} \frac{l!}{(k+2)!(l-k-2)!} f_l q^{l-k-2} +
  \right. \nonumber \\ &&
  \alpha \sum_{m=0}^k \frac{(-1)^{k-m}}{m+1}
  \sum_{l=m}^{\infty} \frac{l!}{m!(l-m)!} f_l q^{l-k-1} +
    \nonumber \\ && \left.
  \alpha \sum_{m=0}^k \frac{(-1)^{k-m}}{k-m+1}
  \sum_{l=m}^{\infty} \frac{l!}{m!(l-m)!} f_l q^{l-k-1-m} \right].
  \end{eqnarray}

Consider the coefficients at the expansion of the function
$F_0(t;q',q)$ in front of the structure
$(it)^n \Delta q^0/ q^{2n+1}$. One can see from~(\ref{f16})
that these coefficients are equal to $-C_{n0}^1$, where
$C_{n0}^1$ is determined so as in the case of the Coulomb
potential by the formula~(\ref{f26}). The contributions arising
from adding of $f(q)$ to the Coulomb potential at~(\ref{f4.1})
change the coefficients in front of the structures
$(it)^n \Delta q^0/ q^m$ only with $m<2n+1$. This means that
factorial growth of $C_{n0}^1$ cannot be cancelled by anything.
Hence, the Schwinger --- DeWitt expansion for the potentials
of type~(\ref{f4.1}), so as for the Coulomb potential, is
divergent. The kernels for such potentials exist only in
asymptotic sense too.

\bigskip

Result of our research is following: it is established that
for the Coulomb and
Coulomb-like potentials the Schwinger --- DeWitt expansion
for the evolution operator kernel is divergent. So, the
kernels for these potentials have additional (beyond $\delta$-like)
singularity at $t=0$. Hence, the initial condition~(\ref{f2})
may be fulfilled only in asymptotic sense. The potentials considered
do not belong to the class of potentials such as~(\ref{f1.3}), for
which the Schwinger --- DeWitt expansion is convergent.

\end{document}